\documentclass{article}
\usepackage{fleqn,ams}
\usepackage{amssymb}
\usepackage{epsfig}
\usepackage{verbatim}

\newcommand{\REFigure}[2]{%
\begin{center}\epsfig{file=#1,width=7cm,height=7cm,angle=-90}\\[12pt]
\refstepcounter{figure}Figure \thefigure: {\sl #2}\end{center}}

\begin{document}

\begin{center}
{\bf\Large  A Kinetics of non-equilibrium Universe. III. Stability of non-equilibrium scenario.}\\
Yu.G.Ignatyev, D.Yu.Ignatyev\\
Kazan State Pedagogical University,\\ Mezhlauk str., 1, Kazan
420021, Russia
\end{center}

\begin{abstract}
An influence of initial distribution nonequilibrium parameters on
thermodynamical equilibrium recovery processes in early Universe
under the assumption of elementary particles' scaling of
interactions in range of superhigh energies is researched.
\end{abstract}

\section{Introduction}
In previous papers of the Authors the model of superheat
particles' cosmological evolution in conditions of scaling of
interactions was built up \cite{LTE, LTE2}. For elementary
particles' cut set of interactions at that the Universal
Asymptotic Cut Set of Scattering, introduced in papers
\cite{UACS,Yuneq}\footnote{system of units $G=\hbar=c=1$.} was
used:
\begin{equation}\label{1}
\Sigma_0(s)=\frac{2\pi}{s\left(1+\ln^2\frac{s}{s_0}\right)}=\frac{2\pi}{s\Lambda(s)},
\end{equation}
where $s$ is a kinematic invariant of four-piece reaction (details
see in \cite{LTE}), $s_0=4$ is a square of two colliding Planck
masses' total energy,
\begin{equation}\label{2}
\Lambda(s)=1+\ln^2\frac{s}{s_0}\approx \mbox{Const},
\end{equation}
In \cite{LTE2} a relaxation of cosmological plasma's superheat component on equilibrium
component was researching under the assumption that particles number in superheat
component is far less than the equilibrium particles number. In particular, it was shown
that the solution of kinetic equation describing an evolution of the
ultra\-re\-la\-ti\-vis\-tic superheat component in the equilibrium cosmological plasma
has a form:
\begin{equation}\label{3}
\Delta f_a(t,\mathcal{P})=\Delta f^0_a(\mathcal{P})
\exp\left[-\frac{\xi(t,\mathcal{P})}{\mathcal{P}}\int\limits_0^t
\frac{y^2(t')dt'}{\sqrt{t'}} \right],
\end{equation}
where $\Delta f^0_a(\mathcal{P})=\Delta f_a(0,\mathcal{P})$
is an initial deviation from equilibrium (detail see in
\cite{LTE2}) and the dimensionless function is introduced:
\begin{equation}\label{4}
y(t)=\frac{T(t)}{T_0(t)};\qquad \sigma\equiv y^4;\qquad
\sigma_0\equiv y^4(0).
\end{equation}
where $T(t)$ is a temperature of plasma's equilibrium component, $T_0(t)$ is a
temperature in the same point of time in completely equilibrium Universe and the
dimensionless momentum variable $\mathcal{P}$ (con\-for\-mal momen\-tum) is introduced:
\begin{equation}\label{5}
p=\mathcal{P}{\cal N}^{1/4}T_0(t).
\end{equation}
Farther $\mathcal{N}$ is an efficient number of particles'
equilibrium types;
\begin{equation}\label{6}
\xi(t,\mathcal{P})=\frac{\pi \tilde{\cal{
N}}}{3\sqrt{\mathcal{N}}}\left(\frac{45}{32\pi^3}\right)^{1/4}%
\frac{1}{\Lambda(\mathcal{P}TT_0/2)};
\end{equation}
is a parameter, weakly dependent on variables $t,\mathcal{P}$,
$$\tilde{{\cal N}}=\frac{1}{2}\left[\sum\limits_B (2S+1)+
\frac{1}{2}\sum\limits_F (2S+1)\right]=N_B+\frac{1}{2}N_F;$$ $N_B$ is a number of sorts
of equilibrium bosons, $N_F$ is a number of sorts of equilibrium fermions. Relative
temperature $y(t)$ of plasma's equilibrium component, presenting a parameter of the
nonequilibrium distri\-bu\-tion (\ref{3}), is determined via the integral equation of
energy-balance:
$$
y^4+\frac{15}{\pi^4}\sum\limits_a (2S_a+1) \int\limits_0^\infty
\mathcal{P}^3\Delta f^0_a(\mathcal{P})\times
$$
\begin{equation}\label{7}
\times\exp\left[-\frac{\xi(t,\mathcal{P})}{\mathcal{P}}\int\limits_0^t\frac{y^2(t')dt'}{\sqrt{t'}}
\right]d \mathcal{P}=1,
\end{equation}
where $(2S_a+1)$ is a statistical factor. From (\ref{7}) in zero
point of time follows the relation \cite{LTE2}:
\begin{equation}\label{8}
\frac{15}{\pi^4}\sum\limits_a (2S_a+1) \int\limits_0^\infty
\mathcal{P}^3\Delta f^0_a(\mathcal{P}) d \mathcal{P}=1-\sigma_0.
\end{equation}
Let us introduce corresponding to \cite{LTE2} a new dimensionless
time variable $\tau$ :
\begin{equation}\label{9}
\tau=\frac{\overline{\xi}}{\overline{\mathcal{P}}_0}\sqrt{t}
\end{equation}
where $\overline{\xi}=\xi(\overline{\mathcal{P}}_0)$;
\begin{equation}\label{11}
\overline{\mathcal{P}}_0=\frac{\sum\limits_a
(2S_a+1)\int\limits_0^\infty d \mathcal{P}\mathcal{P}^3\Delta
f^0_a(\mathcal{P})}
{\sum\limits_a (2S_a+1)\int\limits_0^\infty d
\mathcal{P}\mathcal{P}^2\Delta f^0_a(\mathcal{P})}
\end{equation}
is an average value of momentum variable $\mathcal{P}$ in point of time $t=0$, à òàêæå
áåçðàçìåðíóþ ôóíêöèþ $Z(\tau)$:
\begin{equation}\label{10}
Z(\tau)=2\int\limits_0^\tau y^2(\tau')d\tau'.
\end{equation}
Then subject to (\ref{8}) equation (\ref{7}) is reduced to the
form(\cite{LTE2}):
\begin{equation}\label{12}
y=[1-(1-\sigma_0)\Phi(Z)]^{1/4},
\end{equation}
where
\begin{equation}\label{13}
\Phi(Z)=
\frac{\overline{\mathcal{P}}(t)}{\overline{\mathcal{P}}(0)}=\frac{\sum\limits_a
(2S_a+1)\int\limits_0^\infty d \mathcal{P}\mathcal{P}^3\Delta
f^0_a(\mathcal{P}) e^{-Z\overline{\mathcal{P}}_0/\mathcal{P}}}
{\sum\limits_a (2S_a+1)\int\limits_0^\infty d
\mathcal{P}\mathcal{P}^3\Delta f^0_a(\mathcal{P})}.
\end{equation}%
\section{Numerical Model}
In paper \cite{LTE2} distribution function's initial deviation
from the equilibrium was represented in the most plain form:
\begin{equation}\label{14}
\Delta
f^0(x)=\frac{A}{\mathcal{P}_0^3(k^2+x^2)^{3/2}}\chi(1-x),\quad
k\rightarrow 0,
\end{equation}
where $\chi(z)$ is a Heaviside function (a staircase function),
$x=\mathcal{P}/\mathcal{P}_0$ is a dimensionless momentum
variable, $A$, $\mathcal{P}_0$ and $k$ are certain parameters; $k$
parameter is introduced to provide a convergence of distribution
function's all moments in range of momentum's small values. A
distribution function of superheat particles' energy density at
that has a form, close to the spectrum of the so-called {\it white
noise}, when all energy's values are equiprobable up to the
certain critical value $\mathcal{P}_0$, after which distribution
breaks. From the qualitative point of view an evolution of the
initial distribution (\ref{14}) is reduced to the ``corrosion'' of
nonequlibrium distribution spectrum's low-energy part while energy
density's constancy in high-energy part conserves, what did not
allow to make predictions about the form of distribution's tail
area.
In given paper we select a distri\-bu\-tion function's deviation in more realistic form,
allo\-wing to carry out a research of the relaxation process dependence on parameters of
the initial distribution. The paper's main goal at that is a stability's research of the
nonequlibrium cosmological scenario \cite{LTE2} with respect to the parameters of the
superheat particles' initial distribution. So, let us specify the initial distribution in
form:
\begin{equation}\label{15}
\Delta f^0(\mathcal{P})=Ae^{-\alpha\mathcal{P}},
\end{equation}
so that:
\begin{equation}\label{16}
\Delta\tilde{N}=\sum\limits_a\frac{1}{\pi^2}\int\limits_0^\infty
\Delta f^0(\mathcal{P})\mathcal{P}^2 d\mathcal{P}=\frac{2A}{\pi^2
\alpha^3}\mathcal{N} -
\end{equation}
is an initial conformal density of nonequlibrium particles'
number. Let us calculate an initial conformal energy density of
nonequilibrium particles:
\begin{equation}\label{17}
\tilde{\varepsilon}_1=\sum\limits_a\frac{1}{\pi^2}\int\limits_0^\infty\Delta
f_0{\mathcal{P}\mathcal{P}^3d\mathcal{P}}=\frac{3!A}{\pi^2
\alpha^4}\mathcal{N}.
\end{equation}
Thus, we obtain for the initial average conformal energy:
\begin{equation}\label{18}
\overline{\mathcal{P}}_0=\frac{\Delta\tilde{\varepsilon}}{\Delta\tilde{N}}=\frac{3}{\alpha}.
\end{equation}
Let us calculate the parameter $\sigma_0=\varepsilon_0/(\varepsilon_0+\varepsilon_1)$,
where $\varepsilon_0$ is an energy density of plasma's equilibrium component and
$\varepsilon_1$ is a nonequlibrium component's energy density. Plasma's equilibrium
component is descri\-bed via the distribution:
\begin{equation}\label{19}f_0=B\exp\left(-\frac{p}{T}\right)=
B\exp\left(-\frac{\mathcal{P}\mathcal{N}^{1/4}T_0}{T}\right)=
B\exp\left(-\frac{\mathcal{P}\mathcal{N}^{1/4}}{y}\right).
\end{equation}
Calculating conformal particles number densities and theirs
energies relative to this distribution, we obtain:
\begin{equation}\label{21}
\tilde{\mathcal{N}}_0=\frac{y^3B2!}{\pi^2}\mathcal{N}^{1/4}; \quad
\tilde{\varepsilon}_0=\frac{y^4B 3!}{\pi^2}.
\end{equation}
Comparing (\ref{21}) with an energy density of equilibrium plasma,
which is determined via its temperature by means of the relation
(see \cite{LTE}):
\begin{equation}\label{22}
\varepsilon_0={\cal N} \frac{\pi^2 T^4}{15},
\end{equation}
we find:
\begin{equation}\label{23}
B=\frac{\pi^4}{90}\mathcal{N}.
\end{equation}
Then:\begin{equation}\label{24}
 \mathcal{N}_0=\frac{\pi^2}{45}y^3 \mathcal{N}^{5/4}.
 \end{equation}
Using (\ref{17}), (\ref{21}) and (\ref{23}) we obtain:
\begin{equation}\label{25}
A=(1-\sigma_0)\frac{\pi^4 \alpha^4}{90}.
\end{equation}
Then according to the referred above nonequilibrium particles'
condition of smallness:
 \begin{equation}\label{26}
 n_1(t)\ll n_0(t),
 \end{equation}
it has to be:
\begin{equation}\label{27}
\alpha(1-\sigma_0) \ll y_0^3N^{1/4}.
\end{equation}
On Fig.\ref{7}-\ref{9} the graphs of the initial staircase and expo\-nen\-tial
distributions at equal energy densities and partic\-les' average energies for various
nonequlibrium parame\-ters are shown in comparison.

\REFigure{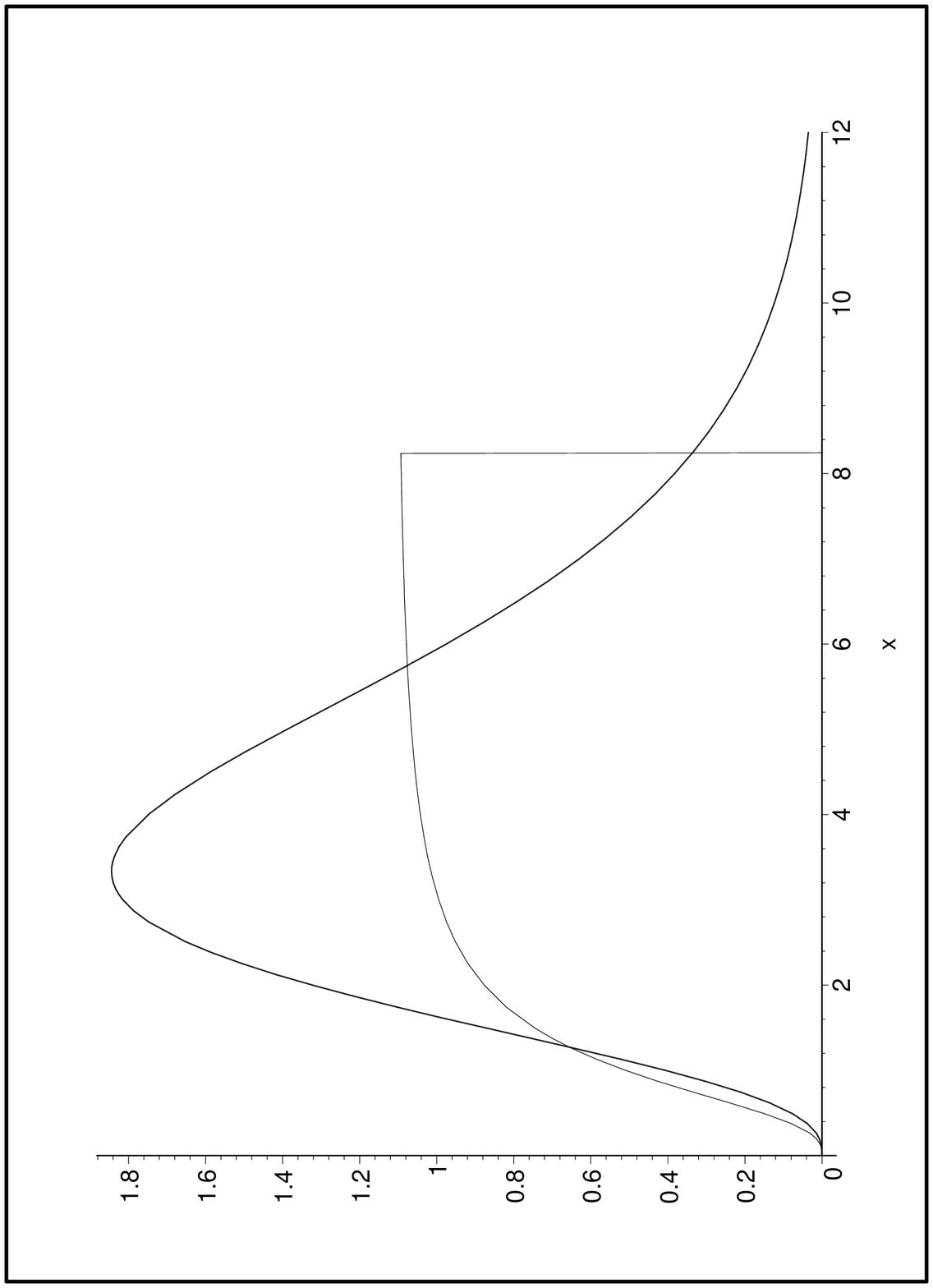}{Comparison of staircase and expo\-nen\-tial
initial distributions at equal in both cases energy densities
$\tilde{\varepsilon}_1$ and average energies of particles
$\overline{\mathcal{P}}_0$; $\alpha=0.9$. Staircase distribution
is a thin line, exponential distribution is a heavy line.
\label{Ris7}}

\REFigure{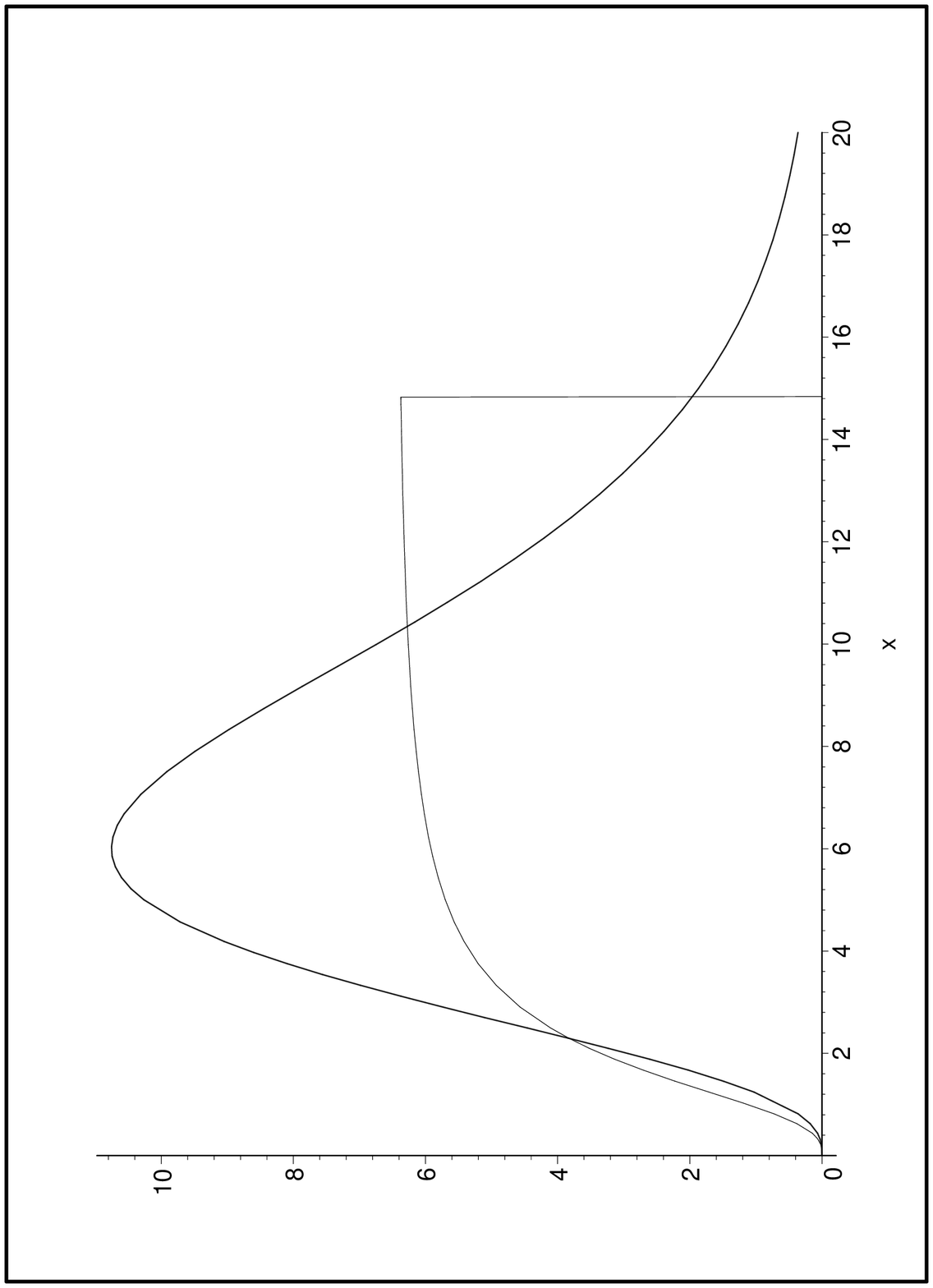}{Comparison of staircase and exponential
initial distributions at equal in both cases energy densities
$\tilde{\varepsilon}_1$ and average energies of particles
$\overline{\mathcal{P}}_0$; $\alpha=0.5$. Staircase distribution
is a thin line, exponential distribution is a heavy line.
\label{Ris8}}

\REFigure{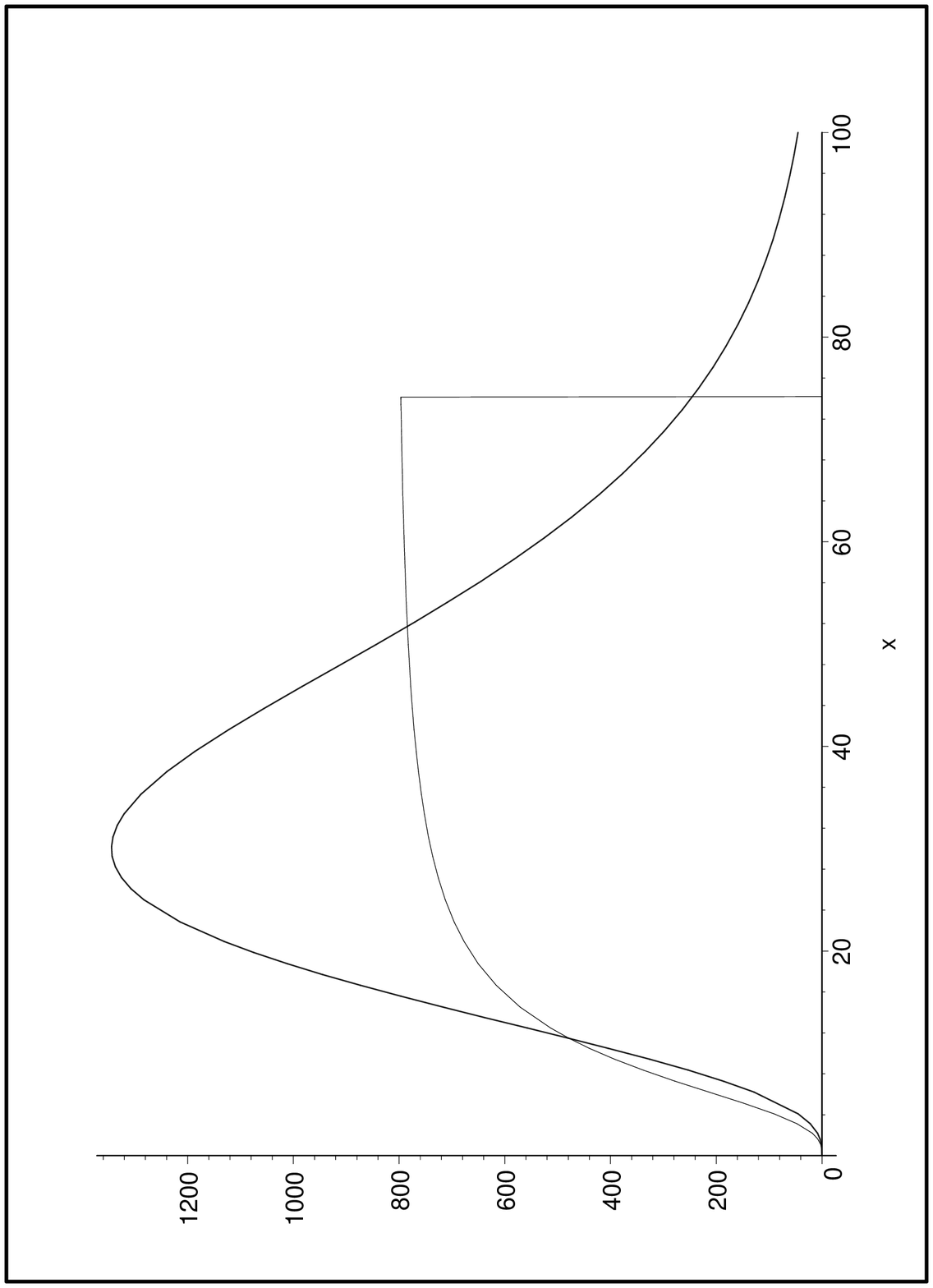}{Comparison of staircase and exponential
initial distributions at equal in both cases energy densities
$\tilde{\varepsilon}_1$ and average energies of particles
$\overline{\mathcal{P}}_0$; $\alpha=0.1$. Staircase distribution
is a thin line, exponential distribution is a heavy line.
\label{Ris9}}

\section{Research of nonequilibrium distribution's relaxation}

Calculating function $\Phi(Z)$ relative to the distribution
(\ref{15}) according to the formula (\ref{13}), we obtain:
$$\Phi(Z)=\frac{9Z^3}{2}\left(\frac{2(3Z+6)K_0(2\sqrt{3Z})}{9Z^2}\right.+$$
\begin{equation}\label{28}+
\left.\frac{4\sqrt{3}(3+6Z)K_1(2\sqrt{3Z})}{27Z^{5/2}}\right),
\end{equation}
where ${K_\nu}(z)$ is a Bessel function of the second sort
(McDonald dunction) \cite{Lebed}:
\begin{eqnarray}\label{29}{K_\nu}(z)=
\frac{\sqrt{n}z^y}{2^y\Gamma(\nu+\frac{1}{2})}
\int\limits_0^\infty{e^{-z\cosh{t}}{\sinh^{2\nu}{tdt}}},\nonumber\\
\Re(z)>0, \Re(\nu)>-\frac{1}{2} .
\end{eqnarray}
Relation of $\tau$ and $Z$ is determined via the following formula \footnote{details see
in \cite{LTE2}.}
\begin{equation}\label{30}
\frac{1}{2}\int\limits_0^Z {\displaystyle
\frac{dU}{\sqrt{1-(1-\sigma_0)\Phi(U)}}}=\tau,
\end{equation}
in which an obtained value of $\Phi(Z)$ (\ref{28})\\ is to be
substituted. On Fig. \ref{Ris1}-\ref{Ris4} the graphs showing an
influence of initial distribution's parameters on plasma's
temperature relaxation process are represented.

\REFigure{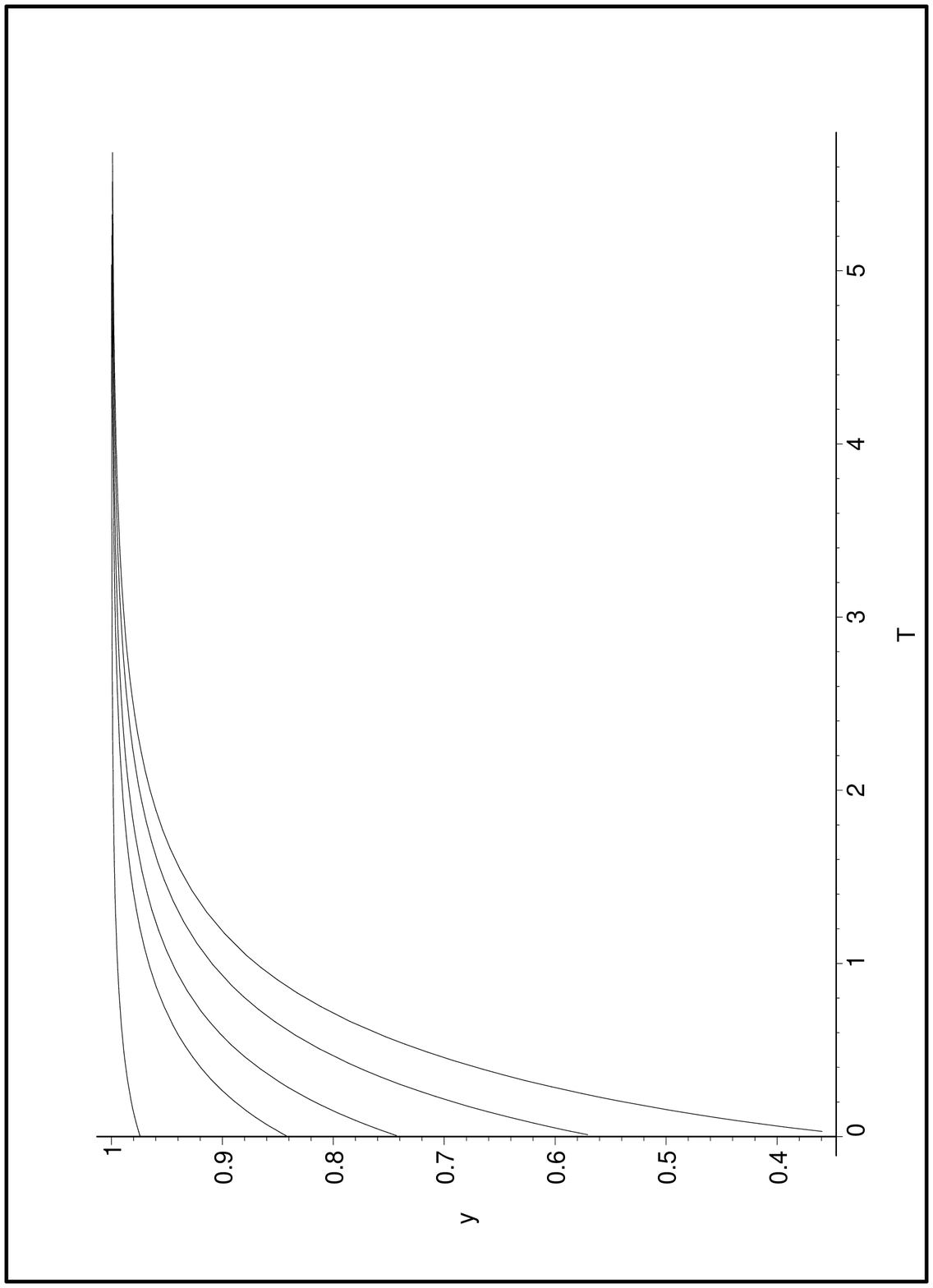}{\label{Ris1}Relaxation of plasma's
temperature to the equilibrium: $y=T(\tau)/T_0(\tau)$ subject to
the parameter $\sigma_0$: – bottom-up $\sigma_0= 0,01;\; 0,1;\;
0,3;\; 0,5;\; 0,9$. Values of dimensionless time variable $\tau$
are put on the abscissa axis.}
\REFigure{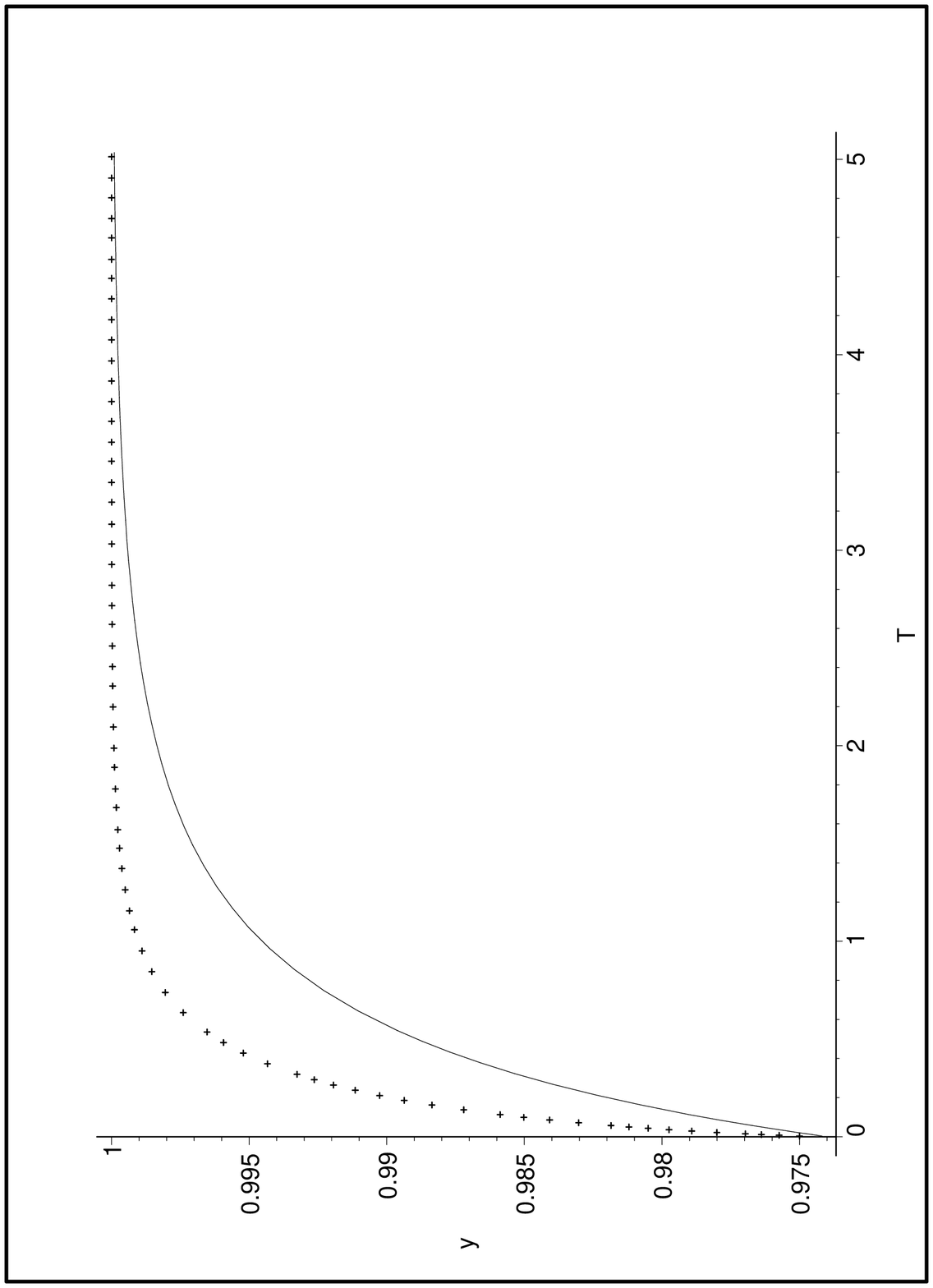}{\label{Ris2}Relaxation of plasma's
temperature to the equilibrium: $y=T(\tau)/T_0(\tau)$ for the
parameter $\sigma_0=0.9$ for staircase function (dotted line) and
$Ae^{-\alpha p}$ function (firm line). Values of dimensionless
time variable $\tau$ are put on the abscissa axis.}
\REFigure{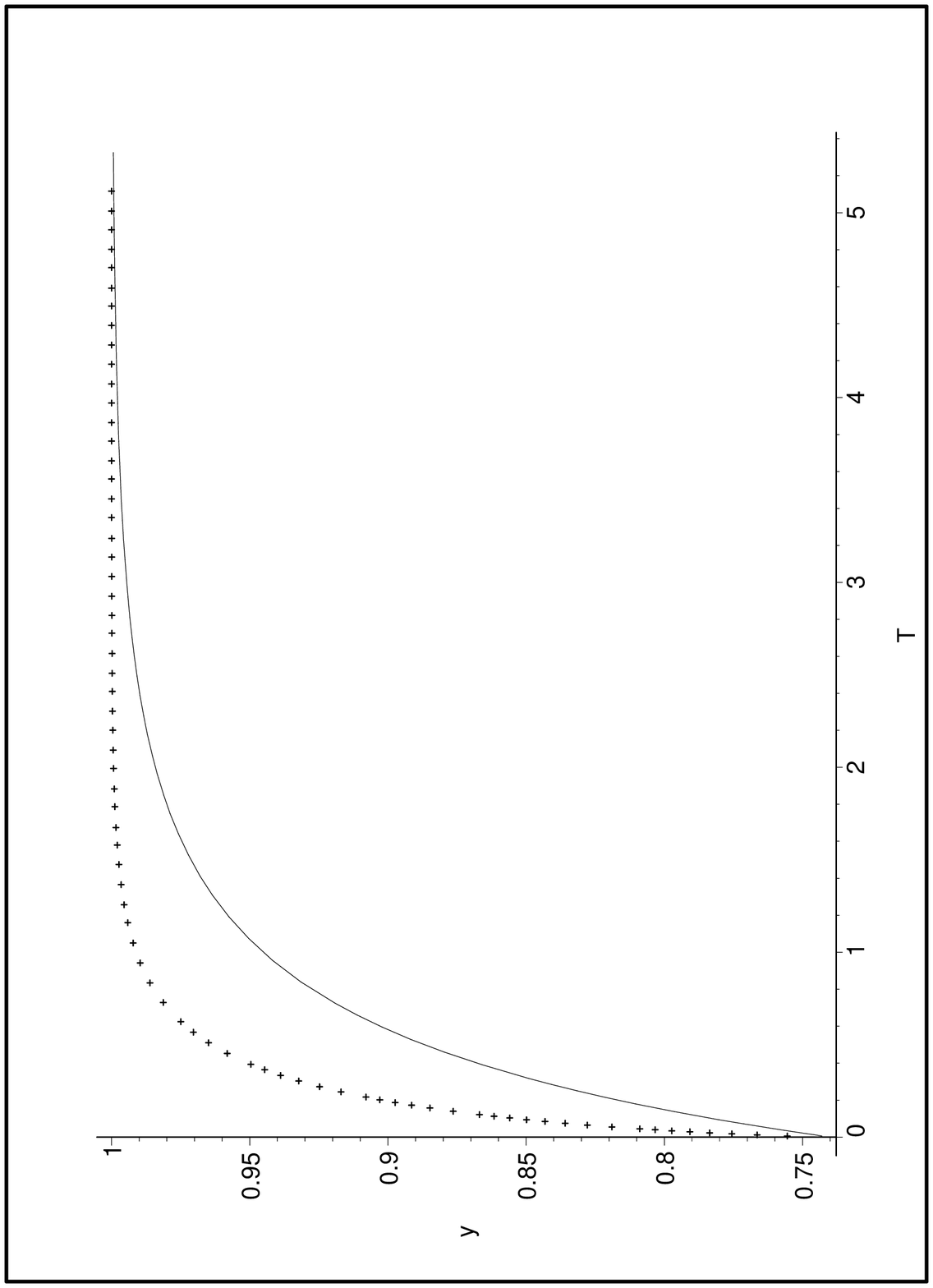}{\label{Ris3}Relaxation of plasma's
temperature to the equilibrium: $y=T(\tau)/T_0(\tau)$ for the
parameter $\sigma_0=0.3$ for staircase function (dotted line) and
$Ae^{-\alpha p}$ function (firm line). Values of dimensionless
time variable $\tau$ are put on the abscissa axis.}
\REFigure{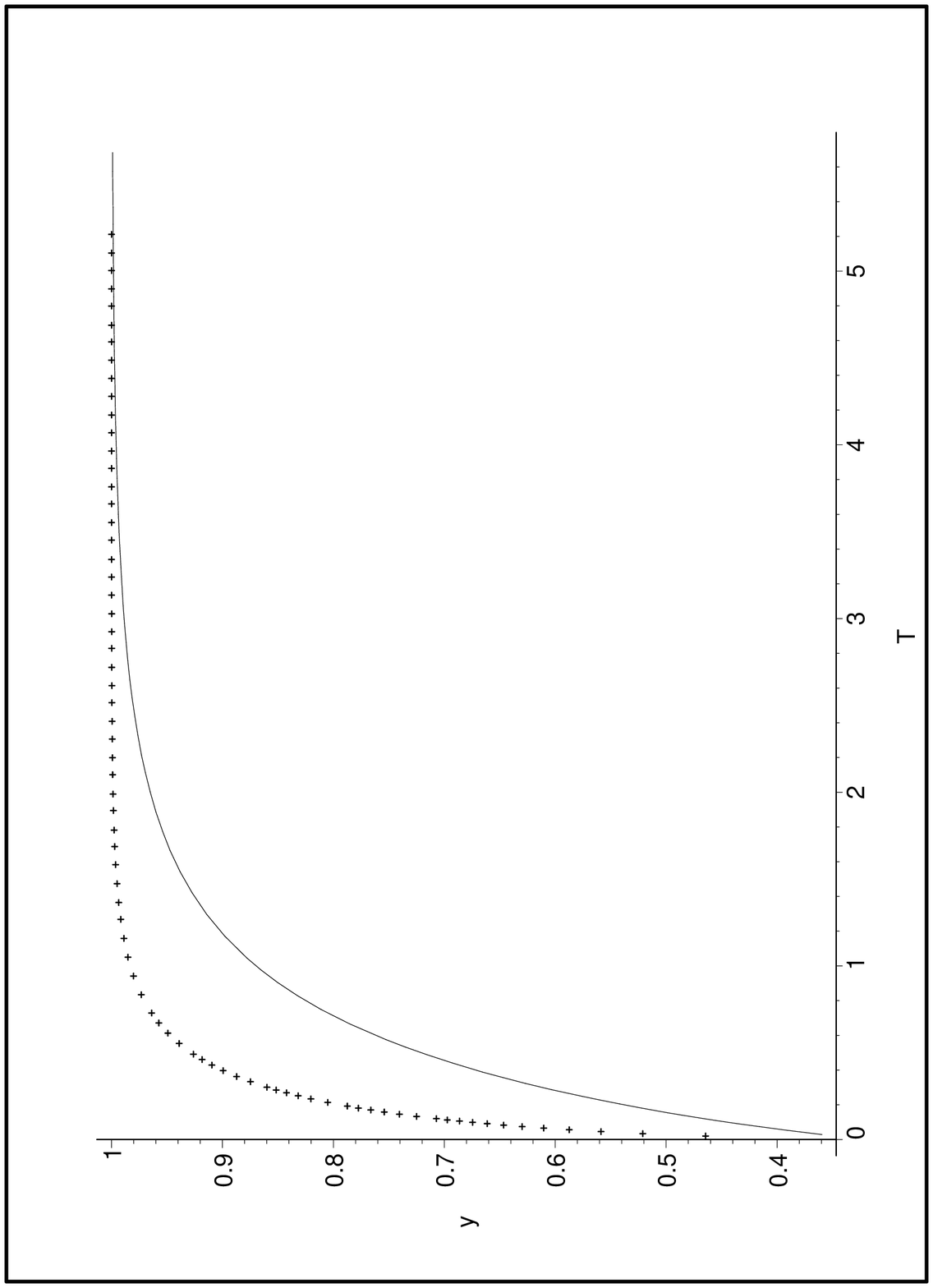}{\label{Ris4}Relaxation of plasma's
temperature to the equilibrium: $y=T(\tau)/T_0(\tau)$ for the
parameter $\sigma_0=0.01$ for staircase function (dotted line) and
$Ae^{-\alpha p}$ function (firm line). Values of dimensionless
time variable $\tau$ are put on the abscissa axis.}

On Fig. \ref{Ris1} the results of numerical modelling of equilibrium component's heating
process by superheat particles for the initial distri\-bu\-tion (\ref{15}) in terms of
equation (\ref{30}) subject to the initial distri\-bu\-tion's para\-me\-ters are shown.
It is obvious from this picture that temperature of
plasma's equilibrium component is satu\-ra\-ted up to the value $T_0(\tau)$ at $\tau\sim 3-4$.%
On Fig.\ref{Ris2}-\ref{Ris4} temperature's relaxation process of the staircase and
exponential distribu\-tions for various values of initial distri\-bu\-tion's
nonequi\-lib\-rium parameter $\sigma_0$ is shown. It is clear from these pictures, that
in the case of the exponential distribution temperature relaxes to the equilibrium rather
slower than in the case of the staircase function, however qualitative behavior of
$y(\tau)$ functions's graphs coincides with each other.

Using the results of equation (\ref{30}) numerical integra\-tion in formula for the
relaxation of the nonequi\-lib\-rium distribution (\ref{3}), we obtain time dependence of
distribution function's deviation from the equilibrium. On Fig.\ref{Ris5} the results of
equations' (\ref{30}) and (\ref{3}) simul\-ta\-neous numerical integration subject to the
value of initial distribution's nonequi\-lib\-rium parameter are shown.

Thus, an evolution of distribution in time variable $Z$ does not depend explicitly from
the nonequi\-lib\-rium parameter $\sigma_0$, but the relation of real time $\tau$ and
time variable $Z$ do. This relation is given by formula (\ref{30}) and is represented in
graphical form on Fig.\ref{Ris5}.

\REFigure{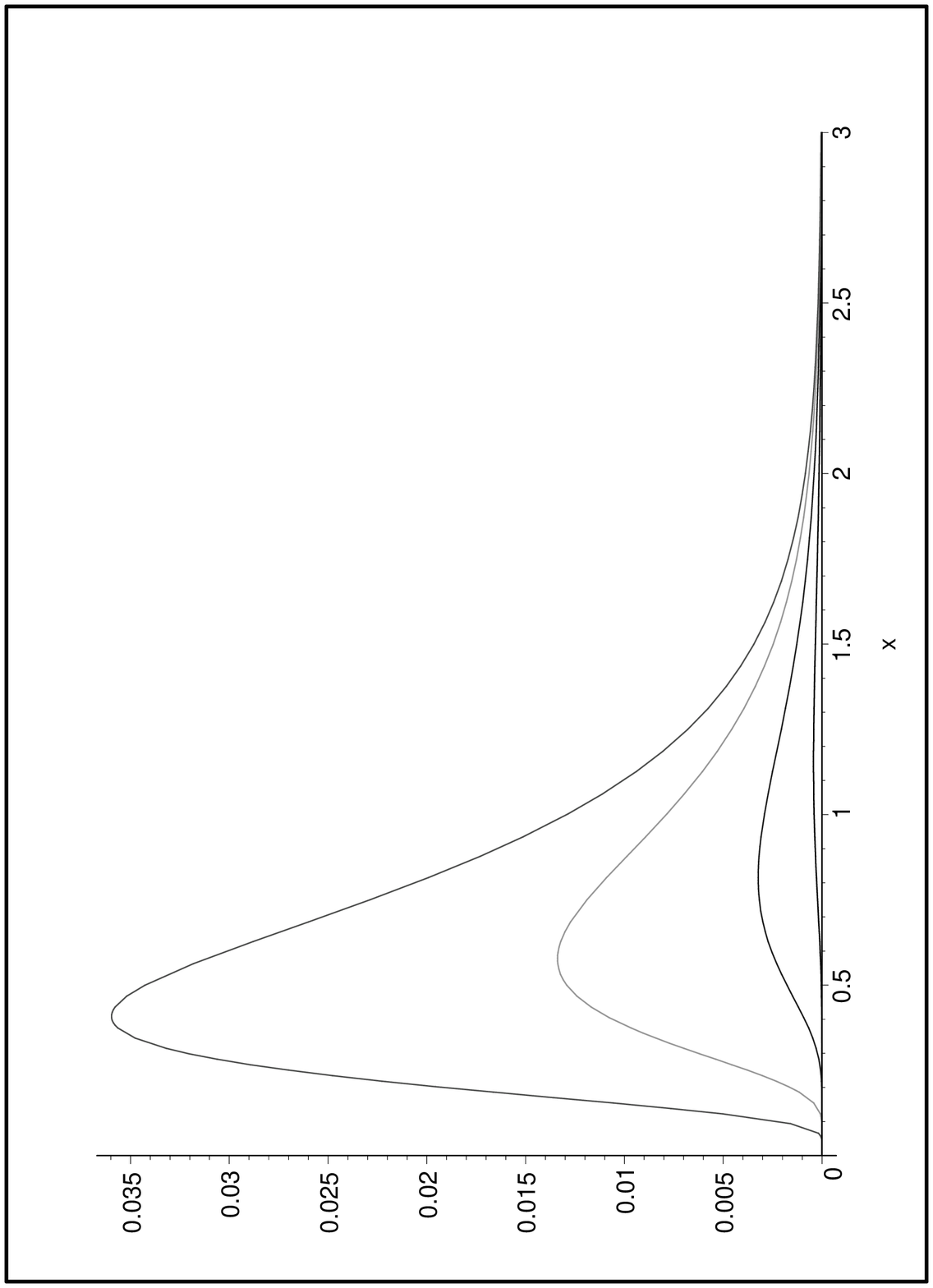}{Relaxation of superheat component for
distribution (\ref{15}) ($A=1$) subject to time parameter $Z$:
top-down - $Z=0.5$, $Z=1$, $Z=2$, $Z=4$.Value of dimensionless
time variable $x$ is put on the abscissa axis, value $\lg(1+f)$ is
put on the ordinate axis. \label{Ris5}}

\REFigure{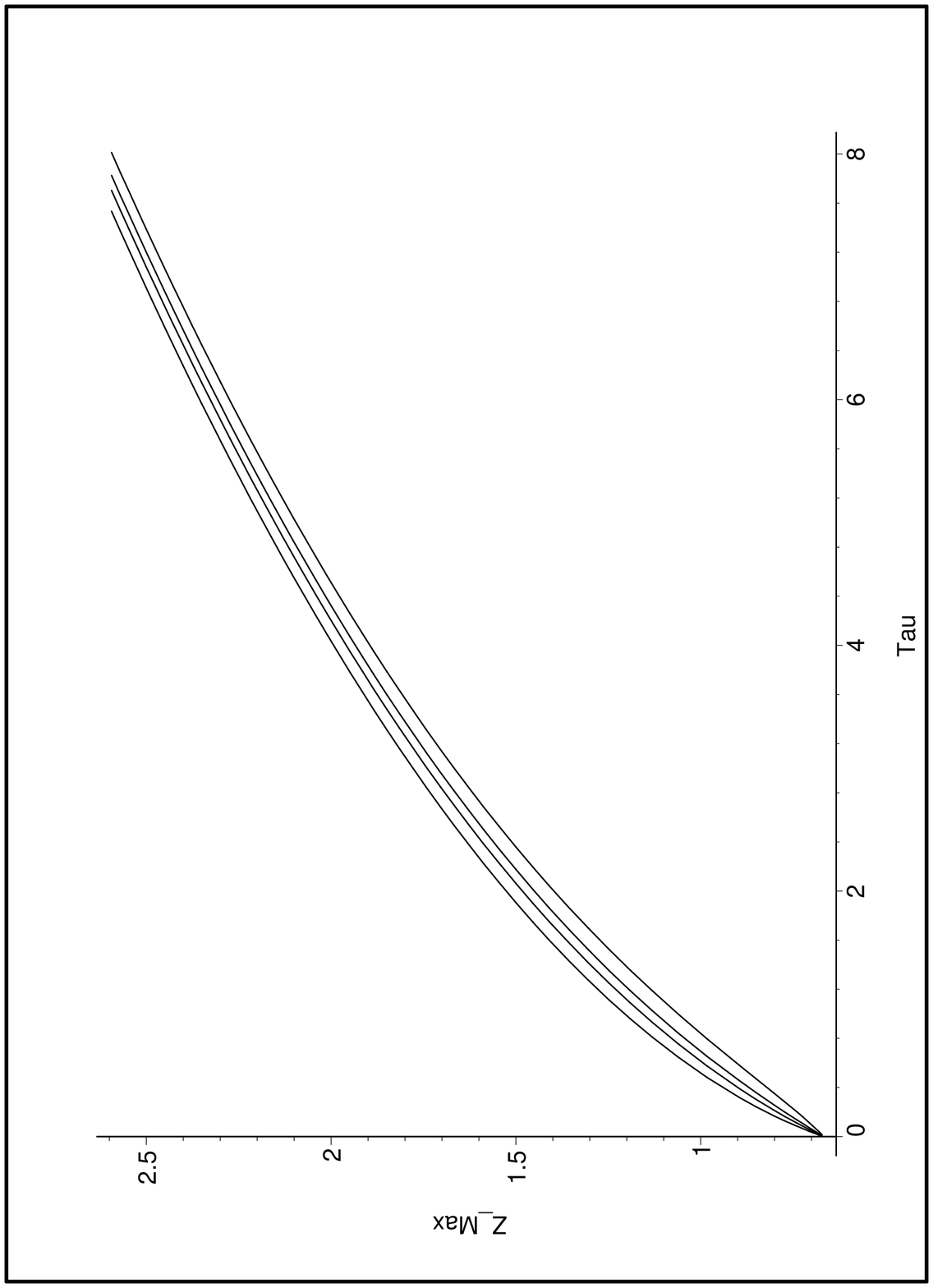}{Dependence of distribution's maximum on
dimensionless time $\tau(Z)$, calculated by formula (\ref{3})
relative to the initial distribution (\ref{15}) subject to initial
distribution's nonequilibrium parameter $\sigma_0$. Top-down:
$\sigma_0=0.01$, $\sigma_0=0.1$, $\sigma_0=0.3$, $\sigma_0=5$,
$\sigma_0=0.9$,
\label{Ris6}} %
\section{Conclusion}
Thus, carried out research has shown from the one hand though
certain differences the stability of nonequilibrium distribution's
relaxation scenario relative to parameters of the superheat
particles' initial distribution and from the other hand has
allowed to reveal the dependence of initial distribution's average
parameters on Universe's evolution time. As evident from
aforecited pictures, maximum of nonequilibrium particles' energy
spectrum shifts with time by the approximate law:
$$\mathcal{P}_{max}\simeq \sqrt{\tau}.$$

In conclusion, Authors express their gratitude to professors V.N.Melnikov,
Yu.S.Vladimirov and D.V.Galt\-sov for the stimulating discussion of paper's matters.

\end{document}